\documentclass[prl,twocolumn,groupedaddress]{revtex4}
\usepackage{graphics,epsfig}

\begin{document}

\title{Optically controlled spin-glasses in multi-qubit cavity systems}
\author{Timothy C. Jarrett, Chiu Fan Lee and Neil F. Johnson}

\affiliation{Physics Department and Center for Quantum Computation, Oxford University, Oxford, OX1 3PU, U.K.}

\date{\today}

\begin{abstract}
Recent advances in nanostructure fabrication and optical control, suggest that it will soon be possible to prepare collections of interacting two-level systems  (i.e. qubits) within an optical cavity. Here we show theoretically that such systems could exhibit novel phase transition phenomena involving spin-glass phases. By contrast with traditional realizations using magnetic solids, these phase transition phenomena are associated with {\em both} matter {\em and} radiation subsystems.  Moreover the various phase transitions should be tunable simply by varying the matter-radiation coupling strength.

\vskip0.2in
\end{abstract}

\maketitle

\newpage
Condensed matter physicists are keen to identify experimental realizations of Ising-like Hamiltonians involving populations of interacting two-level objects (i.e. spins) \cite{Kad}. Exotic phases such as a spin-glass are of particular interest \cite{Sherr}. Experimental studies of spin-glasses have focused on solids containing arrays of magnetic ions \cite{Kad}. However the laws of Nature limit the range of exotic
behaviors that such solids can exhibit, since it is very hard to engineer the
magnitude, anisotropy, range and/or disorder of the spin-spin interaction in
such systems. In the seemingly unrelated fields of atomic, nanostructure and optical physics, there have been rapid advances in the
fabrication and manipulation of effective two-level systems (more commonly referred to as qubits) using atoms, semiconductor quantum dots, and superconducting nanostructures \cite{NatQED,review,QDcavity,ima,books,QD}. In particular, controlled qubit-cavity coupling has been demonstrated experimentally between such systems and a surrounding optical cavity  \cite{NatQED,review,QDcavity,ima,books,QD,PBG}.  In quantum dots systems, in particular, the effective spin-spin (i.e. qubit-qubit) interaction can in principle be tailored by adjusting the quantum dots' size, shape, separation, orientation and the background electrostatic screening.
Furthermore, the interaction's anisotropy can be engineered by choosing asymmetric dot
shapes. Disorder in the qubit-qubit interactions will arise naturally for
self-assembled dots, or can be introduced artificially by varying
the individual dot positions during fabrication \cite{QD,jopa}. 

Motivated by these recent experimental advances, we study the phase transitions which could arise in such multi-qubit-cavity systems. We uncover novel realizations of spin-glasses \cite{Sherr} in which the phase
transition phenomena are associated with the spin (i.e. matter) and boson
(i.e. photon) subsystems. The resulting phase diagrams can be explored experimentally 
by varying the qubit-cavity coupling strength $\lambda$, e.g. by re-positioning the center of the cavity around the nanostructure array, 
changing its orientation, or tuning the cavity Q-factor  \cite{NatQED}.
In addition to opening up the study of these important condensed matter
systems to the nano-optical community, our results help strengthen the
theoretical connection which seems to be emerging between multi-qubit-cavity
systems and spin-spin systems \cite{CMP,CFL}. 

\begin{figure}[ht]
\includegraphics[width=0.5\textwidth]{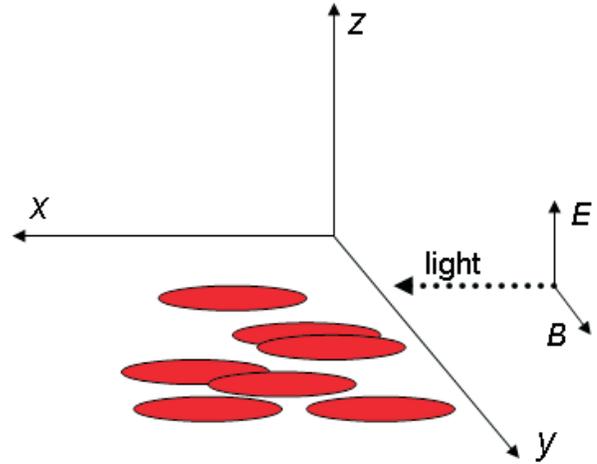}
\caption{
(color online) Schematic diagram showing a possible optical realization of a spin-glass system 
within a multi-qubit-cavity system comprising elongated quantum dots. The quantum 
dots are orientated along the $x$-axis and have non-uniform spacings, giving rise 
to disorder in the qubit-qubit interaction $\{J_{ij}\}$.}
\end{figure}

Given the above experimental considerations, we will introduce a generalization of the well-known Dicke model \cite{CMP,CFL,Dic54,WH73} in order to describe $N$ {\em interacting} two-level systems in a cavity field:
{\small
\begin{eqnarray} 
 H &=& a^\dag a + H_I
     + \frac{\epsilon}{2}\sum_{i=1}^{N}\sigma_{i}^{z}
     + \frac{\lambda}{2\sqrt{N}}\sum_{i=1}^N(\sigma_{i}^{+} + \sigma_{i}^{-})(a + a^\dag), \nonumber
\end{eqnarray}
}
where the operators $a, a^\dag$ and $\sigma^{\pm}_j, \sigma^Z_j$ correspond to 
the photon field and quantum dot $j$ respectively. The interaction term $H_I = \sum_{i<j}J_{ij}\sigma_{i}^{z}\sigma_{j}^{z}$ or
$H_I = \sum_{i<j}J_{ij}\sigma_{i}^{x}\sigma_{j}^{x}$ (or equivalently $\sum_{i<j}J_{ij}\sigma_{i}^{y}\sigma_{j}^{y}$).
An interaction in the $z$-direction is present naturally \cite{review}. 
In the case of the $x$-directed interaction, each quantum dot 
can be engineered to have an elongated form along the $x$-direction, by biasing the
growth process along this direction.  Applying an electric field along $x$, will
then create large dipole moments in that direction. One can use
undoped dots, in which case the dipole results from the exciton, or doped dots,
in which case the dipole originates from the conduction-subband electron biased
along $x$.  A schematic plot of a possible realization is depicted in Fig. 1.
Kiraz et al. \cite{QDcavity}, for example, have already built physical
realizations of such multi-qubit-cavity nanostructure systems.  In particular,
Ref. \cite{QDcavity} shows a scanning electron micrograph of the GaAs microdisk
nanostructure system comprising a disordered array of quantum dots embedded
within an optical cavity.  The cavity has a diameter of $4.5\mu m$ and produced
Q-factor values exceeding 1800  \cite{QDcavity}. The cavity was built with a
collection of InAs quantum dots at irregular locations fixed during the growth, and hence
necessarily features disordered dot-dot couplings $\{J_{ij}\}$.  Another suitable
experimental set-up has been provided by Imamoglu et al.  \cite{ima} in which a
quantized cavity mode and applied laser fields are used to mediate the
interaction between spins of distant, doped quantum dots. This leads to
cavity-assisted spin-flip Raman transitions and hence pairs of quantum dots can
be coupled via virtual photons in the common vacuum cavity mode \cite{ima,review}.

The dot-dot coupling terms $\{J_{ij}\}$ will have an
inherent disorder in self-assembled dots -- alternatively, such disorder can be built in during growth by
varying the dot-dot separations. We make the reasonable assumption 
\cite{Sherr} that the disorder will be Gaussian:
\begin{displaymath}
P(J_{ij})=\left(\frac{N}{2 \pi J^2}\right)^{\frac{1}{2}}{\rm exp}\left(-\frac{J_{ij}^2 N}{2
    J^2}\right)\ \ .
\end{displaymath}
First we consider $H_I = \sum_{i<j}J_{ij}\sigma_{i}^{z}\sigma_{j}^{z}$, noting that the same method can then be used to solve for spin--spin interactions in the other directions.
To obtain the thermodynamical properties of the system, we introduce the
Glauber coherent states  $|\alpha\rangle$ of the field \cite{Glauber}
where 
$a|\alpha \rangle = \alpha | \alpha \rangle$, 
 $\langle \alpha | a^\dag = \langle \alpha | \alpha^*$.
The coherent states are complete,
$\int \frac{d^2\alpha}{\pi}  |\alpha \rangle
\langle \alpha| =1$.
In this basis, we may write the canonical partition function as:
\begin{displaymath} 
Z(N,T)=\sum_{\bf s} \int 
\frac{d^2\alpha}{\pi} 
\langle {\bf s} | \langle \alpha| e^{-\beta H} | \alpha \rangle | {\bf s} \rangle .
\end{displaymath}
As in Ref. \cite{WH73}, we adopt the following assumptions:
\begin{enumerate}
\item
$a/\sqrt{N}$ and $a^\dag/ \sqrt{N}$ exist as $N \rightarrow \infty$;
\item
$\lim_{N \rightarrow
\infty} \lim_{R \rightarrow \infty} \sum_{r=0}^R \frac{(-\beta H_N)^r}{r!}$ can
be interchanged.
\end{enumerate}
 We then obtain:
\begin{eqnarray*} 
 Z &=& {\rm Tr \ } e^{-\beta H} \\
   &=& \int \frac{d^2 \alpha}{\pi} {\rm Tr \exp}\bigg\{-\beta\Big(
         {\rm Re}(\alpha)^2 + {\rm Im}(\alpha)^2 \\
   & &   + \sum_{i<j}J_{ij}\sigma_{i}^{z}\sigma_{j}^{z} 
         + \frac{\epsilon}{2}\sum_{i=1}^{N}\sigma_{i}^{z} 
         + \frac{2\lambda}{\sqrt{N}}{\rm Re}(\alpha)\sum_{i=1}^{N}\sigma_{i}^{x}\Big)\bigg\}.
\end{eqnarray*}
Performing the Gaussian integration of ${\rm Im}(\alpha)$ and setting $w={\rm Re}(\alpha)$
yields the partition function:
{\small
\begin{eqnarray*}
 Z &=& {\rm Tr \exp}\bigg\{-\beta\Big(
     w^2 + \sum_{i<j}J_{ij}\sigma_{i}^{z}\sigma_{j}^{z}
     + \frac{\epsilon}{2}\sum_{i=1}^{N}\sigma_{i}^{z} \\
   & & \qquad \qquad \qquad + \frac{2 \lambda w}{\sqrt{N}}\sum_{i=1}^{N}\sigma_{i}^{x}\Big)\bigg\} .
\end{eqnarray*}
}
We set $J_{ij} \rightarrow -J_{ij}$, $\epsilon = -2\mu$ and $\lambda = -\frac{\Gamma}{2}$
for convenience and rescale the photon field $a^\dag \rightarrow \sqrt{N} b^\dag$
such that $w \rightarrow \sqrt{N} W$.  We then use the Trotter-Suzuki method 
\cite{Thirumalai}, and obtain the free energy from the replica trick:
\begin{displaymath}
-\beta F = [\ln(Z)] = \lim_{n\rightarrow0}\frac{[Z^n]-1}{n}
\end{displaymath}
where $[Z^n]$ is the configurational average of the $n$ replicated partition
functions.  Rewriting the trace and using the method of steepest 
descents, we can proceed in the usual manner for spin-glass problems.  In the thermodynamic limit $N \to \infty$, with 
the limit $n \to 0$ with $N$ kept large but finite, we obtain the free energy,
$f = \frac{F}{N}$, as:
{\small
\begin{eqnarray*}
-\beta f &=& \lim_{\shortstack{{\tiny $n \to 0$} \\ {\tiny $P \to \infty$} \\
                                                    {\tiny $N \to \infty$}}} 
                                                    \frac{[Z^n] - 1}{Nn}\nonumber \\
         &=& \lim_{\shortstack{{\tiny $n \to 0$} \\ {\tiny $P \to \infty$} \\
                                                    {\tiny $N \to \infty$}}}
             \Biggr\{-\frac{1}{2nP^2}\sum_{\alpha,t,t'}(x_{tt'}^{\alpha\alpha})^2 
      - \frac{1}{2nP^2}\sum_{(\alpha,\beta),t,t'}(y_{tt'}^{\alpha\beta})^2 \nonumber \\
& & + \frac{1}{n}\ln\biggl\{{\rm Tr \exp}\Bigl\{
        \frac{\beta J}{\sqrt{2} P^2}\sum_{t,t'}\sum_{(\alpha,\beta)}y_{tt'}^{\alpha\beta}\sigma_{t}^{\alpha}\sigma_{t'}^{\beta} \nonumber \\
& & + \frac{\beta J}{\sqrt{2} P^2}\sum_{t,t'}\sum_{\alpha}x_{tt'}^{\alpha\alpha}\sigma_{t}^{\alpha}\sigma_{t'}^{\alpha}
      + \frac{\beta \mu}{P}\sum_{t}\sum_{\alpha}\sigma_{t}^{\alpha} \nonumber \\
& & - \frac{\beta}{P}\sum_{t}\sum_{\alpha} (W^\alpha)^2 \\
& & + \sum_{t}\sum_{\alpha} \omega^\alpha \left( C^\alpha + \sigma_{t}^{\alpha}\sigma_{t+1}^{\alpha} \right)
      \Bigr\}\biggr\}\Biggr\},                                       
\end{eqnarray*}
}
with $\alpha$ being the replica index and $t=1,2,3,\cdots,P$ the label for the 
Trotter direction.
We then substitute for the following stationary values:
\begin{eqnarray*}
x_{tt'}^{\alpha\alpha} &=& \frac{1}{\sqrt{2}} \beta J \langle \sigma_{\alpha}^t
  \sigma_{\alpha}^{t'} \rangle \equiv \frac{1}{\sqrt{2}}\beta J \chi_{tt'}^{\alpha\alpha} \nonumber \\
y_{tt'}^{\alpha \beta} &=& \frac{1}{\sqrt{2}} \beta J \langle \sigma_{\alpha}^t
  \sigma_{\beta}^{t'} \rangle \equiv \frac{1}{\sqrt{2}}\beta J q_{tt'}^{\alpha\beta} .
\end{eqnarray*}
The replica 
symmetric assumption \cite{Thirumalai} yields
$\chi_{tt'}^{\alpha\alpha}=\chi_{tt'}$, 
$q_{tt'}^{\alpha\beta}=q_{tt'}$, 
$W^\alpha=W$, and with the static approximation we then obtain
$\chi_{tt'}=\chi$ and  $q_{tt'}=q$. The
validity of this approximation has been discussed at length in Ref. \cite{Thirumalai}.  Hence:
\begin{widetext}
{\small
\begin{eqnarray*}
-\beta f &=& \lim_{\shortstack{{\tiny $n \to 0$} \\ {\tiny $P \to \infty$} \\
                                                    {\tiny $N \to \infty$}}}
             \Biggl\{-\beta W^2 -\frac{\beta^2 J^2 \chi^2}{4} 
       - \frac{\beta^2 J^2 (n-1) q^2}{4}
       + \frac{1}{n}\ln \left\{ \int Dz_1 \prod_{\alpha} \int Dz_{2_\alpha}
        {\rm e}^{\left(\sum_{t}\omega C \right)}
        {\rm Tr \ e}^{\left(
        \sum_{t}\left(h_\alpha \sigma_{\alpha}^{t} + \omega \sigma_{\alpha}^{t}\sigma_{\alpha}^{t+1}\right)
        \right)}\right\}\Biggr\}
\end{eqnarray*}
}
\end{widetext}
with
{\small
\begin{displaymath}
h_\alpha = \frac{\beta J}{P} \left( \sqrt{q}z_1+\sqrt{\chi-q}z_{2_\alpha}+\frac{\mu}{J}\right),
\end{displaymath}
}
{\small$\int Dz_1 = \sqrt{\frac{1}{2 \pi}} \int \exp \left\{ -\frac{z_1^2}{2}
\right\}$} and {\small$\int Dz_{2_\alpha} = \sqrt{\frac{1}{2 \pi}} \int \exp \left\{ -\frac{z_{2_\alpha}^2}{2}
\right\}$}.  We then calculate the trace and perform power series expansions in the limit 
$P \to \infty$.  Finally, we expand out the last term in the free energy, take the limit $n \to 0$
and replace $\mu$ and $\Gamma$ by their original values $\mu =
-\frac{\epsilon}{2}$ and $\Gamma = -2\lambda$ to obtain:
{\small
\begin{eqnarray*}
-\beta f &=&  - \beta W^2 -\frac{\beta^2 J^2}{4}\left(\chi^2 - q^2 \right) \\  
     & & + \int Dz_1 \ln\left(\int Dz_{2} 2 \cosh\Omega \right)
\end{eqnarray*}
}
with
\begin{eqnarray}
\Omega &=& \left(b^2 + \beta^2 \lambda^2 W^2\right)^\frac{1}{2}, \nonumber \\
b &=& \beta J \left( \sqrt{q}z_1+\sqrt{\chi-q}z_{2}-\frac{\epsilon}{J}\right),
\nonumber
\end{eqnarray}
having absorbed the factors $\frac{1}{2}$ and $4$ into $\epsilon$ and
$\lambda$ respectively.
The requirement that the free energy is stationary yields the following 
self-consistent equations:
{\small
\begin{eqnarray*}
\chi &=& \int Dz_1 \left\{ \frac{\int Dz_2 \left\{ \left(
    \frac{b^2}{\Omega^2}\right) \cosh \Omega + \left(\frac{\beta^2 \lambda^2 W^2}{\Omega ^3}\right)
    \sinh \Omega \right\}}
    {\int Dz_2 \cosh \Omega} \right\} , \\
q &=& \int Dz_1 \left( \frac{ \int Dz_2 \left\{ \left(\frac{b}{\Omega}\right) \sinh \Omega
    \right\}}{\int Dz_2 \cosh \Omega} \right)^2 , \\
W &=& \frac{1}{2} \int Dz_1 \left\{ \frac{ \int Dz_2 \left( \sinh \Omega 
    \left(\frac{\beta \lambda^2 W}{\Omega}\right)\right)}{\int Dz_2 \cosh \Omega} \right\}.
\end{eqnarray*}
}
Adding $\Delta H = h \sigma_i^z$ to the Hamiltonian such
that $\epsilon \to (\epsilon + h)$ and taking the limit $h \to 0$, yields:
{\small
\begin{displaymath}
m = -\frac{1}{\beta} \frac{\partial f}{\partial h}\bigg\vert_{h=0} 
  = - \int Dz_1 \left\{ \frac{\int Dz_2 \left\{ \left( \frac{b}{\Omega} \right) \sinh \Omega \right\}}
      {\int Dz_2 \cosh \Omega} \right\}.
\end{displaymath}
}

\begin{figure}[ht]
\includegraphics[width=0.5\textwidth]{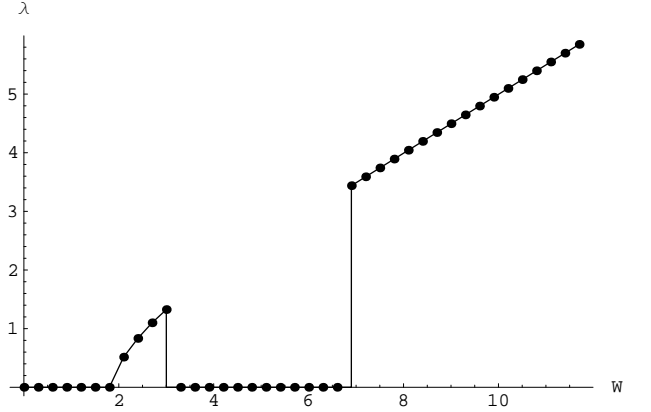}
\caption{A plot of the magnitude of the super-radiant order parameter, $W$,
against $\lambda$ with $\beta=2$, $\epsilon=1$ and $J=2$, and a spin--spin
interaction in the $z$-direction.  The gap between the two non-zero $W$ phases
disappears as we decrease the temperature, $T$.
}
\end{figure}
\begin{figure}[ht]
\includegraphics[width=0.5\textwidth]{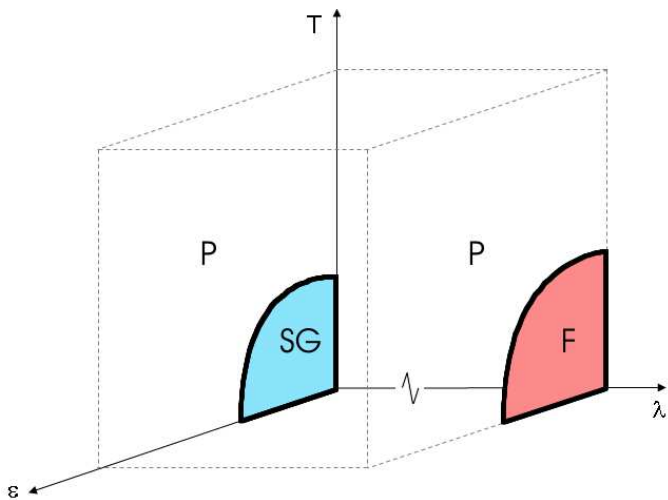}
\caption{(color online) Schematic plot of the phase diagram of our system when a spin--spin
coupling is present in the $x$-direction.  As we increase $\lambda$, the phase
diagram changes from one which resembles that  of an Ising spin-glass model 
with a transverse field, to one with only ferromagnetic and 
paramagnetic phases (i.e. the spin-glass phase has disappeared).
}
\end{figure}

An intricate array of phase transitions arises in {\em both} the matter {\em and} the optical sub-systems. Furthermore, these transitions in the matter and optical sub-systems are inter-dependent.  Here we summarize 
the main results for a general interaction $H_I =\sum_{i<j}J_{ij}\sigma_{i}^{z}\sigma_{j}^{z}+
\sum_{i<j}J_{ij}\sigma_{i}^{x}\sigma_{j}^{x}$.  
If the interaction is only in 
the $z$-direction, a ferromagnetic-paramagnetic phase transition emerges in
the matter sub-system -- but there is no spin-glass phase anywhere in the phase diagram.
This is to be expected, as at any finite value of $\epsilon$ there is a non-zero
contribution to the magnetisation as $\epsilon$ acts as a
longitudinal term to the Ising interaction \cite{Sherr}.  As such, we would only expect a
spin glass phase in the limit of $\epsilon \to 0$.
However, novel behavior emerges in the optical sub-system in the form of  {\em two}
sub-superradiant phase transitions, as shown in Fig. 2.  This
effect disappears as we increase $\beta$.  We see that the system goes through a first order and then a second order
phase transition as we increase $\lambda$ \cite{CFL}.  {\it In particular, the sharp nature of both 
transitions and the narrowing of the gap between them as $\beta$ increases, suggests that 
an accurate low-temperature thermometer could be constructed based on this
effect.}

The addition of the $x$-directional interaction to the original Dicke Hamiltonian yields a number of
interesting results.  In particular, the condition for the system to
exhibit a sub-superradiant phase transition is drastically changed from the case without interactions.  If $J$ is
large, the matter sub-system swamps the optical sub-system and the
system is unable to reach superradiance.
Consequently the magnetization is always zero.  However, lowering $J$ yields a superradiant phase -- and by varying $\epsilon$ we uncover a spin-glass phase that moves into a paramagnetic phase at higher $\epsilon$.
Most remarkably, there are {\em two} different phase diagrams for the system depending on the
value of the qubit-cavity coupling $\lambda$ at a given $J$.  The corresponding phase boundaries are indicated in Fig. 3.  The complex behavior in the region between these two limits will be discussed in depth elsewhere. 

To summarize, our work shows that novel phase transition phenomena will arise in suitable optical 
realizations of generalized spin-spin Hamiltonians. 
In addition to their intrinsic theoretical interest, we hope that our 
findings might motivate experimental efforts toward exploring these predicted phases.

\end{document}